# Vehicle-to-Everything: Looking into the Future of Flexibility Services

Omid Homaee[a], Chaimaa Essayeh[a], Vahid Vahidinasab[a,*], Ville Tikka[b], Gonçalo Mendes[c] and Jim Fawcett[d]

[a]Department of Engineering, Nottingham Trent University, Nottingham, United Kingdom.
[b]LUT Business School, Yliopistonkatu 34, Lappeenranta, 53850, Finland.
[c]LUT School of Energy Systems, Yliopistonkatu 34, Lappeenranta, 53850, Finland.
[d]Future Isle of Wight CIC, United Kingdom.

### Abstract

The primary aim of this paper is to illuminate potential Vehicle-to-Everything (V2X) flexibility services that can be activated at the three levels of home, community, and grid. To do this, the potential practical services that can be provided by EVs, flexibility requesters, and the required exchange mechanisms at these three levels are identified. At the home level, the two main services that EVs can provide to households are explored. The initial service focuses on cost reduction for homes by employing smart charging and discharging methods, and the second service underscores the capability of an EV equipped with bidirectional chargers to function as a backup resource during grid outages. At the community level, three flexibility services are introduced and outlined: community cost reduction, energy sharing, and backup resources. There is more than one flexibility requester, including the community manager, EV owners, and other end users. Accordingly, at this level, having a fair and transparent market mechanism can optimise the activation of different V2X flexibility services. At the grid level, flexibility providers can offer two main flexibility services, namely load profile adjustment and real-time voltage and frequency control, to a wide range of flexibility requesters, including distribution network/systems operators (DNOs/DSOs) to overcome technical challenges of the distribution network, energy suppliers to manage their energy portfolio, and TSOs to support transmission network. In addition, a review of trial V2X projects and V2X supporting regulations is presented to contextualize the feasibility and regulatory framework for implementing these services. This analysis offers insights into the challenges and opportunities that need to be addressed for the effective integration of V2X flexibility services across the three identified levels.

Word count: 9954

## Contents

*Vahid.vahidinasab@ntu.ac.uk
ORCID(s): 0000-0002-0779-8727 (V. Vahidinasab);
0000-0002-6395-3373 (V. Tikka)

## Abbreviations

| | |
|---|---|
| AI | Artificial Intelligence |
| BRP | Balance Responsible Party |
| BDCH | Bi-Directional Charger |
| BEMS | Building Energy Management System |
| CS | Charging station |
| CCL | Community Cost Lowering |
| DR | Demand Response |
| DCH | Unidirectional Charger |
| DNO | Distribution Network Operator |
| DRES | Distributed Renewable Energy Sources |
| EV | Electric Vehicle |
| EVSE | Electric Vehicle Supply Equipment |
| FSP | Flexibility Service Provider |
| HEMS | Home Energy Management System |
| HCL | Home Cost Lowering |
| MILP | Mixed-Integer Linear Programming |
| SOC | State of Charge |
| TSO | Transmission System Operator |
| V1B | Smart charging at building |
| V1H | Smart charging at home |
| V1C | Smart charging at communities |
| V2B | Vehicle to Building |
| V2G | Vehicle to Grid |
| V2H | Vehicle to Home |
| V2X | Vehicle to everything |
| VPPs | Virtual Power Plants |
| PHEV | Plug-in Hybrid Electric Vehicle |
| LMP | Locational Marginal Pricing |

## 1. Introduction

The net-zero target is an initiative taken by a coalition of countries, including the biggest polluters, that aims at cutting carbon emissions to a small amount of residual emissions that can be absorbed and durably stored by nature [1]. This initiative comes as a reaction to the increasing concerns of global warming and responds to the urgency of addressing climate change and the need for ambitious mitigation efforts to limit global warming [2].

The transportation sector contributes around 25% of global carbon dioxide ($CO_2$) emissions [3], and its decarbonisation plays a pivotal role in achieving the net-zero emissions target. Among the various strategies for reducing emissions from transportation, electrification presents a key solution [4]. Therefore, there is an aim to expedite the adoption of vehicle electrification and propel the evolution of smart cities across the globe. However, the integration of EVs, particularly on a scale, comes with several challenges that need to be addressed to ensure a smooth transition to net-zero transportation:

- Grid Capacity and Stability [5–7]: The increased demand for electricity from charging EVs can strain the existing grid infrastructure, leading to congestion and potential instability. Upgrading grid capacity is very costly and will reflect directly on the end customer energy bill.

- Peak Load Management [8]: Charging EVs during peak hours can exacerbate peak load demand on the grid, requiring additional generation and grid capacity to meet the surge in electricity demand. New strategies need to be implemented to help manage peak loads and optimize grid utilization.

- Grid Congestion and Distribution Upgrades [9–11]: Concentrated EV charging in certain areas can lead to grid congestion and voltage fluctuations, necessitating distribution upgrades to accommodate the increased demand.

- Grid Resilience and Reliability [12,13]: The distributed nature of EV charging introduces variability and uncertainty into the grid, which can affect grid resilience and reliability.

By implementing smart charging solutions, such as time-of-use pricing and DR programs, we can unlock the full potential of EVs to reduce greenhouse gas emissions, enhance energy security, and drive the transition to a sustainable energy future. At the core of this initiative lies the strategic integration of bi-directional smart charging technologies and the promotion of V2X awareness.

Smart Charging (SC) [14] (sometimes referred to as V1X) refers to the intelligent and optimized management of EV charging processes. V2X on the other hand, is the umbrella term that covers vehicle-to-home (V2H), vehicle-to-building (V2B), and vehicle-to-grid (V2G) [15] services that enable vehicles to discharge electricity in support of the entity connected to it. SC and V2X turn challenges into opportunities by harnessing the potential flexibility of EVs to minimize the impact of EV charging on the electrical grid, maximize the utilization of renewable energy sources, and reduce overall energy costs for EV owners. It should also be noted that since V2X can reduce the need for grid upgrades, all energy users benefit financially whether or not they are EV owners.

When it comes to V2X, battery degradation is a major concern for EV owners. The impacts of V2G on battery degradation have been reviewed in [16]. The primary battery degradation mechanisms that result in capacity and power fade have been identified. Subsequently, various battery degradation models have been examined, highlighting their respective advantages and disadvantages. In [17], a stress-based empirical model was created using data obtained from conducting various tests and experiments. Then, this model has been used to design the remuneration strategy to cover the full cost imposed on EV drivers due to participation in V2G services. In [18], payments required to offset the degradation costs of EVs participating in V2G services have been estimated. Techno-economic analyses in [18–23] have shown that even though providing one V2G service might not be financially viable, providing a combination of V2G services has the potential to be profitable for the EV user, even considering battery degradation.

In [24], it has been stated that V2X-capable EVs may not be of interest to EV buyers at this time. Concerns about limited range, stringent V2G contracts, and expensive batteries are the main factors contributing to this result. Therefore, identifying innovative services that can be provided by V2X-capable EVs can provide a new revenue stream and consequently increase the willingness to pay for these EVs. EV user acceptance for smart charging and utility-controlled charging has been investigated in [25, 26]. The survey on a set of early adopters from Germany in [26] has shown that effectively conveying the advantages of smart charging, such as enhancing grid stability and integrating renewable energy sources, to users has a huge impact on the acceptance rate. On the other hand, users' preference for personalized and adaptable mobility poses a challenge to the acceptance of smart charging. Similar studies have been done to assess the user acceptance of V2G in the Netherlands [27], Germany [28, 29], Canada [25], Switzerland [30, 31], United States [32, 33], South Korea [34], and Norway [35].

Multiple research studies have explored V2X opportunities, offering diverse models and designs for V2X applications. In [36], the general framework of V2H and V2G was presented, along with mathematical modeling of constraints and objectives. Meanwhile, [37] examined technical, social, and regulatory challenges identified in 47 V2X trials, synthesizing insights from interviews with experts involved in these projects and providing implications for policymakers, industry, and academia. Additionally, [38] reviewed various economic frameworks and associated regulations from a market perspective to enable value streams for V2X. A comprehensive overview of the latest advancements in system modeling and optimization techniques for effectively managing the interconnected energy and transportation system has been outlined in [39]. Recent progress in EVs and their infrastructure, focusing on advancements driven by AI have been reviewed in [40]. It critically evaluates the use of AI in enhancing EV performance, streamlining EV charging stations, and integrating EVs into the smart grid. Frequency and ancillary services that can be provided by EVs have been identified and categorized in [41]. However, these studies often adopted a singular perspective on V2X, overlooking the other facets and failing to depict the interactions of different roles at the business layer, neither did they present real-world applications. Building upon the insights gathered from the current literature, this paper delves deeper into the realm of local V2X flexibility services. As a natural progression, our investigation aims to bridge the current gaps by linking different views and offers insights into the landscape of V2X solutions, while aligning with the broader objectives outlined within the DriVe2X project [1]. The main contributions of the paper are:

- Synthesizing different V2X services from both academic and industrial perspectives,

- Identification and classification of the different V2X flexibility services.

- Analysis of different players and exchange mechanisms for the identified services at the business layer.

- Review of trial projects and synthesis of the lessons learned from these projects.

## 2. Home level

In the future, household energy consumption may see a rising trend due to the mass deployment of EVs and consequently increasing the number of home-level EV chargers [42]. Nonetheless, these EVs may also yield certain benefits. Their role in fulfilling the energy requirements towards more sustainable net zero energy buildings has been assessed in [43]. When an EV connected to a household's electrical infrastructure is intended to work under the V2H mode, its charging and discharging schedules need to be controlled to fulfill the household's requirements, either manually or automatically. Two primary services that EVs can offer to households are discussed. The first service involves reducing home costs through smart charging and discharging. The second service highlights the potential of an EV with bidirectional chargers to act as a backup resource for emergency situations when the grid is offline. It should be noted that when it comes to the home level, since the service provider and service requester are the same, there will be no need for an exchange mechanism. Based on [44], there are different strategies, including load curtailment, timed charging, smart metering integration, self-consumption of locally generated renewable energy, and V2H, to charge/discharge EVs at the home level with different hardware and software requirements, challenges, and opportunities. Considering all possible charging and discharging strategies, potential services that an EV can introduce to a household have been illustrated in Fig.1. It can be seen from this figure that a DCH charger can provide Home Cost Lowering (HCL) service to the household through four different V1H charging strategies. On the other hand, BDCH chargers not only can lower the energy cost of a home through these four V1H charging strategies, but they can utilize the V2H strategy to lower the energy cost even more. A categorization of some of the works that have been done is presented in Table 1.

Before elaborating on V1H and V2H charging strategies, we need to define unrestricted charging. This strategy is often referred to as 'dumb charging'. The configuration describes a singular EV charging system retrofitted as an additional circuit to an existing electrical installation. The component layer of this charging strategy is illustrated in Figure 2. This strategy allows EV owners to use electricity for charging without any time or tariff restrictions. The consumption for EV charging is entirely demand-driven, beginning as soon as the vehicle is plugged in and ending when it is fully charged, posing potential challenges for grid capacity during peak times. Users are not restricted in their EV charging behaviour, and high energy costs during peak times are not considered problematic by the consumer [44]. However, in some situations, these high costs may indeed be problematic,

---
[1]https://drive2x.eu/

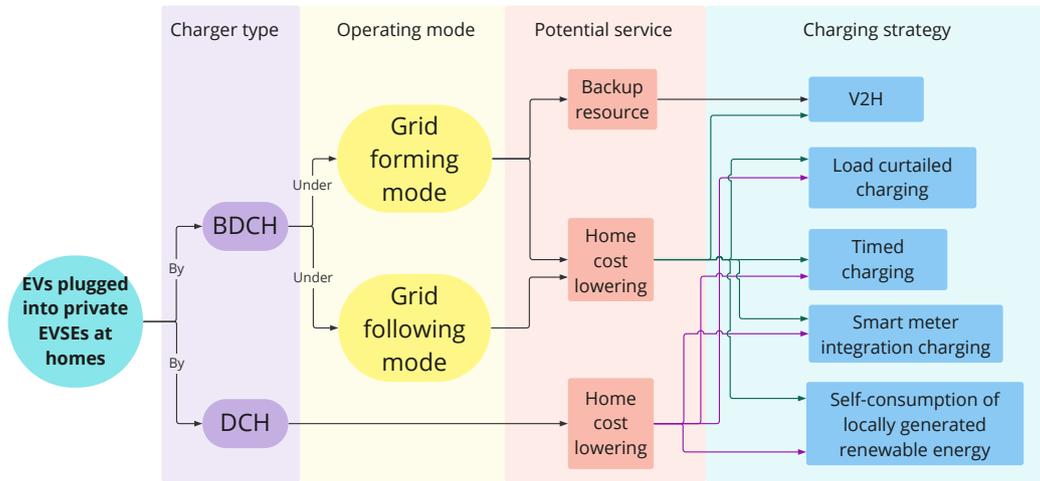

**Figure 1:** Flexibility services at the home level

but users lack the equipment and/or skills to set up a smart charging regime

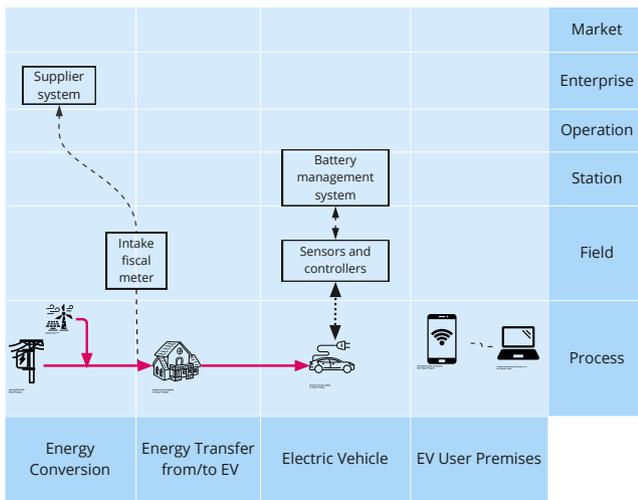

**Figure 2:** Component layer of unrestricted charging

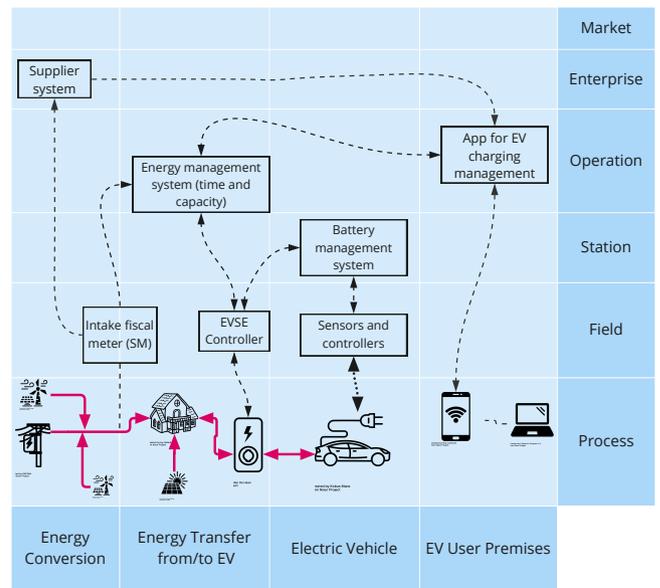

**Figure 3:** Component layer of home cost lowering

## 2.1. Service 1: HCL

HCL involves minimizing the cost of energy usage of the household, including the energy needed to charge the EV and power other household activities. To enable this service effectively, HEMS software is required. With the primary objective of reducing home costs in mind for EV owners, several strategies can be employed to achieve this goal. The component layer is shown in Figure 3. This section is dedicated to the discussion of these strategies. As it can be seen from Fig.1, the first four charging strategies can be used even by DCHs, and the last one can only be used by BDCHs.

### 2.1.1. HCL by of a timed charging control

This charging strategy typically involves restricting EV charger usage to off-peak periods, utilizing basic timer controls. To use this charging strategy, the primary electrical intake needs to be a smart meter to enable the energy supplier to measure the energy consumption during the peak and off-peak timeslots. An override for necessary peak-time charging may also be available. Since there is no need for optimization to obtain the charging and discharging time slots, this method may require manual checks to ensure the energy consumption aligns with the total electrical capacity and other load demands [44]. This strategy is the suitable one for end users engaging in the DR initiative with TOU pricing [47].

### 2.1.2. HCL by load curtailed charging

This charging strategy is designed for monitoring and controlling electrical loads throughout the entire installation. In situations where the total load approaches the maximum capacity, and there is an additional need for EV charging, the integrated energy management controls can assess the capacity and adjust electrical loads accordingly to avoid overload. The controls can deactivate less critical electrical loads to make room for faster EV charging, while crucial loads remain operational. Alternatively, the EV charging power can be reduced to prioritize important loads, extending the charging time. A smart meter needs to be connected to the energy management system, enabling flexible charging power based on the demands of the inflexible electrical installation. This strategy can also be used to manage capacity constraints on the upstream distribution network. To enable this, energy suppliers or DNOs/DSOs may impose restrictions on the home's demand to manage the total demand of feeders. Adjusting the charging power when immediate EV use is not essential can be a preferred option, also offering potential reductions in network connection costs [44].

**Table 1**
Sevices that can be provided by EVs to households

| Reference | EV technology | Service | Charging strategy | Local energy resources | Sell energy to the grid | EV battery degradation modeling |
|---|---|---|---|---|---|---|
| [45] | V2H | Home cost lowering | smart meter integration | PV | Yes | No |
| [46] | V2H | Home cost lowering | timed charging | battery storage | Yes | No |
| [47] | V1H | Home cost lowering | 1) Timed charging, 2)smart meter integration, and 3)load curtailed program | PV and battery storage | Yes | No |
| [48] | V2H | Home cost lowering | Timed charging | - | Yes | No |
| [49] | V2H | Home cost lowering | smart meter integration | - | Yes | Yes |
| [50] | V2H | Home cost lowering | smart meter integration | PV, and battery storage | Yes | Yes |
| [51] | V2H | Home cost lowering | Timed charging | PV | Yes | No |
| [52] | V2H | Home cost lowering | Timed charging and smart meter integration | PV, and battery storage | No | No |
| [53] | V2H | Backup resource | Feeding the household | PV | NA | No |
| [54] | V2H | Backup resource | Feeding the household | PV | NA | No |
| [55] | V2H | Backup resource | Feeding the household | PV | NA | No |
| [56] | V2H | Backup resource (for planned outages) | Feeding the household | PV | NA | No |
| [57] | V2H | Home cost lowering | smart meter integration | PV | No | No |

In this case, the electricity price is more likely to be flat and the EV drivers will be incentivized to use their EVs by incentive-based DR programs such as the V1H model presented in [47]. Also, these load curtailments can be instigated by energy suppliers as part of a DR program [58]. DR strategies, as load shifting tools, are proposed in [59, 60] to limit the peak load by controlling the charging profile of EVs. In [61], a DR program is investigated to constrain the peak power usage of households with V2H-capable EVs to a predetermined value during specific hours. In this study, a MILP model is developed to minimize the total expense of the household's electricity usage by optimizing the scheduling of EV charging and discharging alongside other local energy resources. Although numerical studies indicate a slightly lower cost reduction compared to dynamic pricing scenarios, this approach can be favored due to its ease of implementation.

### 2.1.3. HCL by smart meter integration charging

This charging strategy is designed for monitoring and controlling energy costs, with a specific emphasis on managing the cost of EV charging. Utilizing smart meter-led controls, the primary objective is to facilitate EV charging during periods of the lowest tariffs. In this strategy, typically, EV charger electricity usage is confined to off-peak periods when tariffs are usually lower. In other words, the integration of smart metering allows the charger to automatically adjust to dynamic tariffs that change frequently throughout the day, optimizing cost savings. However, it's important to note that this method may not inherently consider the risks associated with energy supply limitations, potential impact on other loads within the same installation, or potential strain on the upstream infrastructure capacity [44]. This strategy should be used by end users who receive real-time energy prices through their smart meters [47].

### 2.1.4. HCL by self-consumption of locally generated renewable energy

This approach involves leveraging local renewable generation, to reduce dependence on the grid supply. It can also be considered as a strategy for carbon reduction at the household level. It may also include the deployment of local energy storage to optimize energy usage within the home or for EV charging. The strategy aims to utilize renewable energy on-site instead of exporting it, particularly during peak times when tariffs are typically higher [44].

### 2.1.5. HCL by V2H

This approach involves a bi-directional EV charging system. The EV draws energy from the home's electric power infrastructure but also has the potential to provide energy back to it. Full integration with smart metering and local energy management is essential for this setup. This strategy allows the EV to serve as a potential energy storage, storing energy during off-peak times when the energy price is low. This stored energy can be consumed within the home during more expensive tariff periods or when the home's load exceeds the contracted capacity. Energy management is crucial to balance the needs of the EV and the household [44].

Various charging strategies, such as unrestricted charging when the electricity tariff is flat, timed charging when the tariff is either Economy 7 or ToU, and smart meter integration, have been analyzed in [52] for the UK case. The numerical findings indicate that the timed charging strategy exerts the most significant strain on the grid, resulting in higher peaks in demand throughout the week due to reduced electricity costs during specific times of the day. However, from the EV user's point of view, these strategies offer the lowest energy costs for charging. In contrast, the flat tariff yields a more consistent demand pattern as electricity prices remain steady throughout the day. However, it accrues the highest electricity costs among all analyzed tariffs due to the fixed pricing structure throughout the day. The smart meter integration strategy falls between the other strategies, displaying high demand peaks during periods of low prices and lower peaks during times of high prices, reflecting the frequent fluctuations in electricity pricing. Overall, the total electricity cost is slightly higher than that of timed charging control.

## 2.2. Service 2: backup resource

Aside from lowering household energy cost, EVs can also serve as a backup resource for the household. In other words, when a power outage occurs (whether it is an unexpected outage due to a fault or a planned outage due to a maintenance program in the power system), and the household loses power supply, the energy stored in the EV's battery can be utilized to power the emergency or medical appliances in the household. The component layer of this strategy is illustrated in Figure 4. It is important to note that the EV charging system should be able to work under the grid-forming mode to provide this service. It means that the converter should be able to adjust the output voltage and frequency. Moreover, a transfer switch is needed to isolate the household from the distribution system [62]. The strategy to manage the EV to be able to provide this service depends on the type of outage. It should be noted that the reduction of the minimum permissible SOC of the EV to allow it to continue discharging to the house during the outages, will reduce the range of the EV should it be needed immediately after the outage. This is a trade-off that needs to be managed.

Even though optimal charging and discharging of EVs in homes have received significant attention in recent years [57, 63, 64], there is limited research on scheduling them for standalone homes during power outages. An open-source computational tool has been developed in [62] to analyze EV abilities as backup resources during power outages. Based on numerical studies for various home sizes, weather conditions, backup source setups, and load configurations, by limiting consumption to critical loads, the backup duration can be extended on average from 2.7 days to 25 days. [55] has proposed a mixed integer quadratically constrained programming model to maximize the backup duration that a PHEV can provide to a home equipped with rooftop solar panels. In [65], the utilization of EVs as a backup energy source is delineated into two distinct scenarios: firstly, when the EV battery charger provides energy to electrical appliances in isolated systems, and secondly, when it functions as an offline UPS. For the latter case, implementing an algorithm to detect power outages is imperative, whereas such a requirement is unnecessary for the former case. Planning of off-grid smart homes including solar panels, wind turbines, and energy storage systems considering the V2H and V2G technologies has been studied in [66–68].

## 3. Community level

Multiple EVs connected to the electrical circuit of a community (residential and commercial buildings, airports, parking lots, and microgrids, ...) can generate several promising opportunities for the entire energy ecosystem, including the community's electrical network operator, EV owners within the community, and other energy end users in the community. They have the potential to enhance sustainable energy communities [69,70]. These opportunities need to be activated through a fair and transparent exchange mechanism. Figure 5 illustrates different aspects of flexibility services at the community level. As shown in this figure, the EVs connected to the homes inside the community can also be considered as flexibility providers at this level. So, V2X flexibility providers at the community level can be categorized into two categories as follows:

- EVs plugged into grid-connected homes in the community
- EVs plugged into public grid-connected charging stations in the community

### 3.1. Flexibility services at the community level

Table 2 provides a detailed overview of the works done to define services that EVs can offer to communities. In the following these services are elaborated.

#### 3.1.1. Service 1: CCL

CCL can be considered as a service that plugged-in EVs can provide to the community manager. Although the superficial goal of this service may involve community load curtailment during high-demand periods or maximizing self-consumption of locally generated renewable energy when

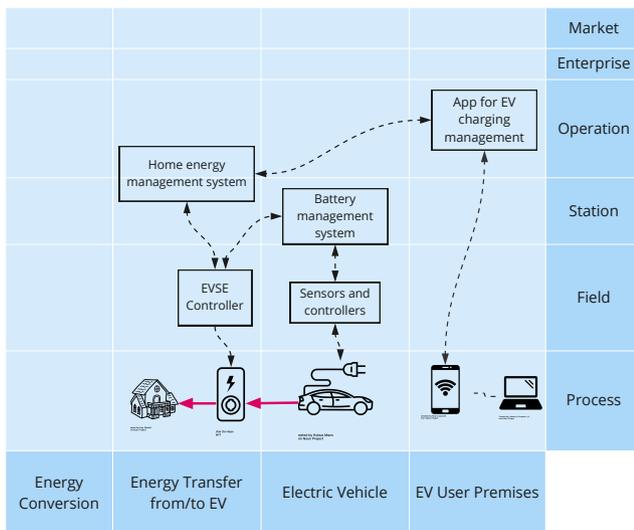

**Figure 4:** Component layer of EVs as the backup resource

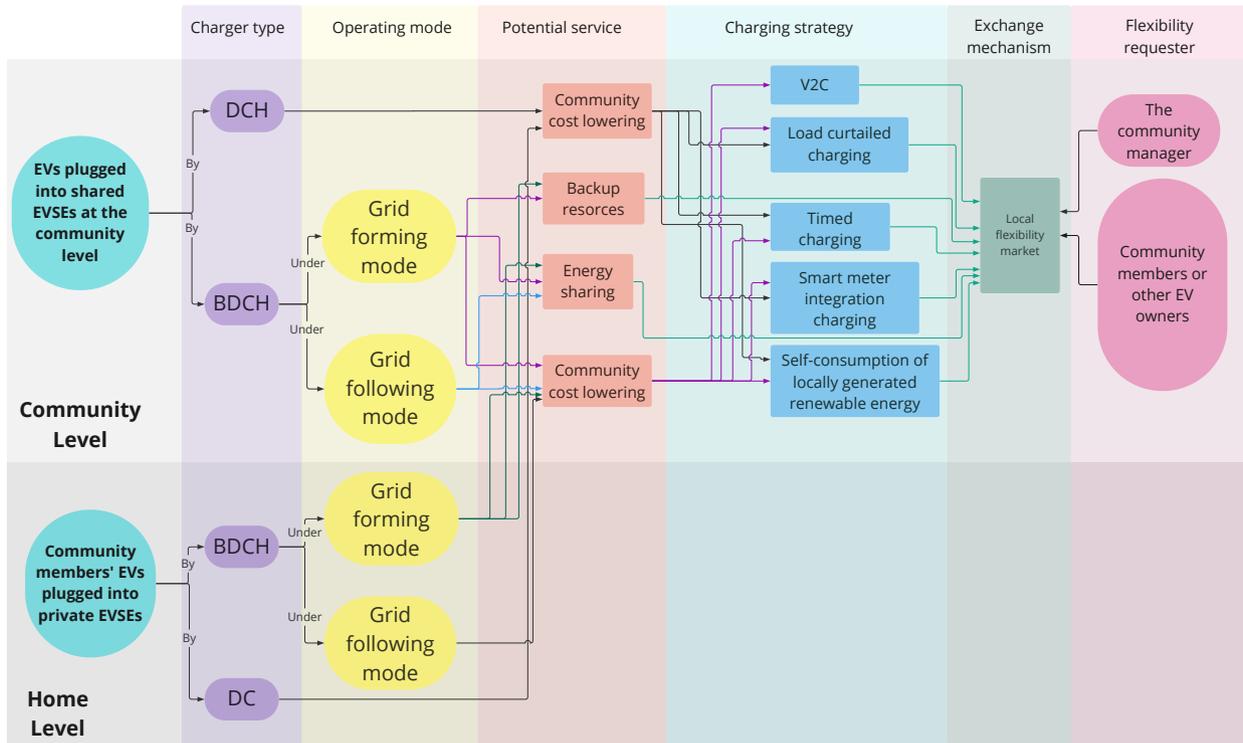

**Figure 5:** Flexibility services at the community level

**Table 2**
Sevices that can be provided by EVs to communities

| Reference | EV technology | Type of the community | Service | Local energy resources | Sell energy to the grid | Flexibility requester | EV battery degradation modeling |
|---|---|---|---|---|---|---|---|
| [71] | V2H | Neighbouring households | Energy sharing | PV, energy storage | Yes | The neighbouring household | Yes |
| [72] | V2C | Apartment building | Community cost lowering | PV, energy storage | Yes | building manager | No |
| [73] | V2C | Apartment building | Community cost lowering | PV, energy storage | Yes | building manager | No |
| [55] | V2C | multiple homes | Backup resource | PV | NA | homes without backup resources | No |
| [74] | V2C | Residential | Backup resource (for planned outages) | PV | NA | Community manager | No |
| [75] | V2B | cluster of buildings | Energy sharing | PV | No | buildings | No |
| [76] | V2C | building | Load leveling and reducing the building's dependence on the electrical grid * | PV, battery storage systems | No | building manager | Yes |
| [77] | V2C | building | Energy sharing | PV, battery storage systems | Yes | other buildings | No |
| [78] | V2B | Building | Community cost lowering | PV | Yes | building manger | No |
| [79] | V2B | buildings | Community cost lowering | PV, and energy storage battery | Yes | | No |

*The building under investigation in this study is the Civil Engineering Building of the National Taiwan University in Taiwan. Given their priority to reduce fluctuations in building energy consumption and less reliance on the power grid, EVs have been utilized to achieve these goals.

production exceeds demand in certain hours, the primary objective from the requester's standpoint is to minimize the cost of energy consumption in the community. To maximize the participation of EV owners in this service, a transparent and fair exchange mechanism is required. In this context, we introduce the concept of the micro flexibility market as the designated exchange platform in the communities. This market serves as a platform where every willing EV owner can offer their EV's battery flexibility for sale to the community manager. It operates as a facilitator, allowing

EV owners to actively participate in the optimization of their EV batteries for the benefit of the community's energy management. In other words, the micro flexibility market at the community level creates a structured platform for EV owners and community managers to engage in a mutually beneficial exchange of flexibility services. In the following, the available strategies that can lead to lowering community costs are presented. It should be noted that a detailed agreement confirming the extent of flexibility that each EV owner is prepared to offer can be used as an easier to implement alternative for the micro flexibility market. Completing the session at a designated SOC will be an important part of this.

1. CCL by timed charging control
   This strategy is similar to HCL by timed charging control. The community manager needs to buy the required flexibility through the exchange mechanism. Subsequently, the community energy management system will gain control over the EVSE controller, replacing the autonomy of EV owners who have sold their flexibility, based on their agreement.

2. CCL by managed charging control
   The primary goal of this strategy is to avoid overload on the community's electrical infrastructure due to the simultaneous charging of several EVs. The electrical intake features a smart meter linked to the community's energy management system. Smart charging operations are controlled by the community manager, who needs to control the overall load on the entire installation. This central balancing ensures that the combined electrical capacity of the installation remains within limits, preventing overloads.

3. CCL by load curtailed charging or by smart meter integration charging
   CCL by load curtailed charging and by smart meter integration charging are similar to HCL by load curtailed charging and by smart meter integration charging, respectively. It should be noted that load curtailed charging can be considered as a generalized form of managed charging control.

4. CCL by self-consumption of locally generated renewable energy
   This strategy is similar to HCL by self-consumption of locally generated renewable energy. Here it is assumed that the ownership of the local energy resources is with the community manager. Accordingly, the energy generated by these local resources can be offered in the micro flexibility market and EV owners can participate to buy it. This arrangement creates a relationship where the community manager can monetize excess energy production, while EV owners can benefit from potentially lower-cost (compared to the network energy cost at the moment), locally sourced energy.

5. CCL by V2C
   This is similar to HCL by V2H. Its component layer is shown in Figure 6. In this strategy, EVs can participate in the micro flexibility market and sell their energy to the community manager. The community manager then buys the required energy to be able to lower the cost of energy usage in the community. For example, EVs can be charged when electricity prices are low and discharged when prices are high. In other words, they are leveraging their capacity as energy storage units. This arrangement can be a win-win for both parties:

the community manager can reduce the community's energy costs, and the EV owners can generate profit.

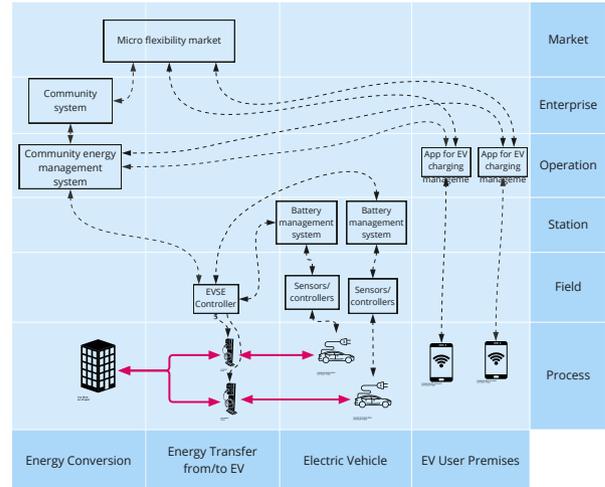

**Figure 6:** Component layer of community cost lowering

### 3.1.2. Service 2: energy sharing

Through this service, EVs can share the energy stored in their batteries with other EVs and end-users within the community. This energy sharing can be facilitated through the micro-flexibility market. The component layer is similar to the component layer of community cost reduction via V2C.

### 3.1.3. Service 3: backup resource

As discussed in the previous section, EVs with V2H capabilities can serve as backup resources for emergency loads. In communities accommodating multiple EVs, this capability can be leveraged as a tradable commodity. In such instances, EV owners with this capability can offer and sell their services as backup resources to requesters within the community. These requesters may include the community manager, other EVs, and other energy end users in the community. Its component layer is illustrated in Figure 7. [55] has proposed a model titled "Vs2Hs" for energy sharing during power outages in communities. One of the reasons for considering Vs2Hs is that they could offer backup power to homes lacking energy sources like PV generators and PHEVs.

## 3.2. Activation of flexibility services at the community level

Activation of V2C services can be accomplished through two approaches. The first approach involves pure competition, where a micro flexibility market is established. The second approach utilizes the community manager as the facilitator. The business layer of this approach is illustrated in Figure 8. As illustrated in this figure, the community manager purchases energy from a supplier and sells it to EV owners, all the while providing parking spaces for their EVs. EV owners as the flexibility providers can participate in the micro flexibility market while the community manager, EV owners, and other energy end users in the community can be considered as flexibility requesters in this micro market. This approach has the potential to activate all three V2B services listed earlier in this section. However, there is a simpler alternative approach through the community manager as the facilitator. In this approach, the community manager as the only flexibility buyer is the price maker and the EV owner has to choose whether to participate or not. It is worth mentioning that the activation with this approach will only be successful if it gives the EV owner adequate control, particularly SOC, and effectively monetizes EVs' participation as well as compensate for battery degradation. The business layer of this structure is illustrated in Figure 9. It should be noted that in this approach the participation rate of EV owners is expected to be less than the case with pure competition.

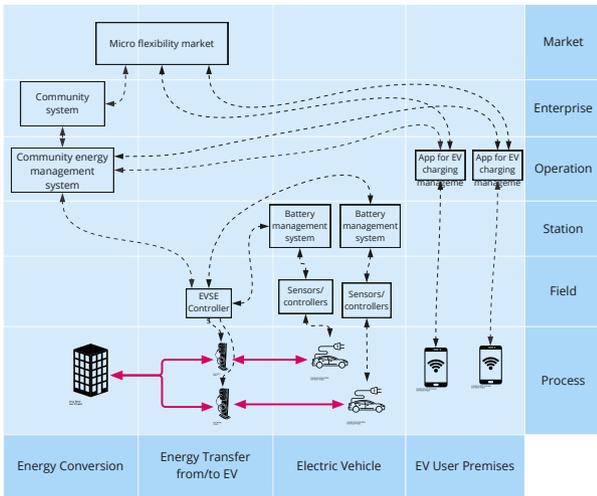

**Figure 7:** Component layer of backup resource at the community level

## 4. Grid level

EVs can present both challenges and opportunities from the grid perspective. The effects of EV integration on distribution networks have been reviewed from various viewpoints in [80–84]. This section is intended to provide an overview of the flexibility services at the grid level. This includes the identification of flexibility providers, flexibility requesters, and potential flexibility services. Figure 10 illustrates an overview of these services at the system level.

### 4.1. V2X flexibility providers at the grid level

At the grid level, all EVs connected to the power system, including the distribution network, are potential providers of flexibility. These EVs can be classified into three categories:

- EVs plugged into grid-connected homes
- EVs plugged into grid-connected communities
- EVs connected to public charging stations

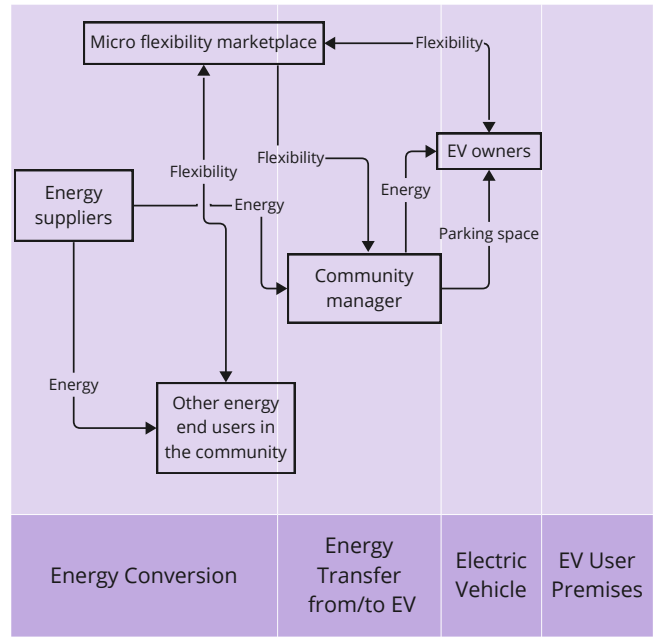

**Figure 8:** Business layer of V2C with pure competition

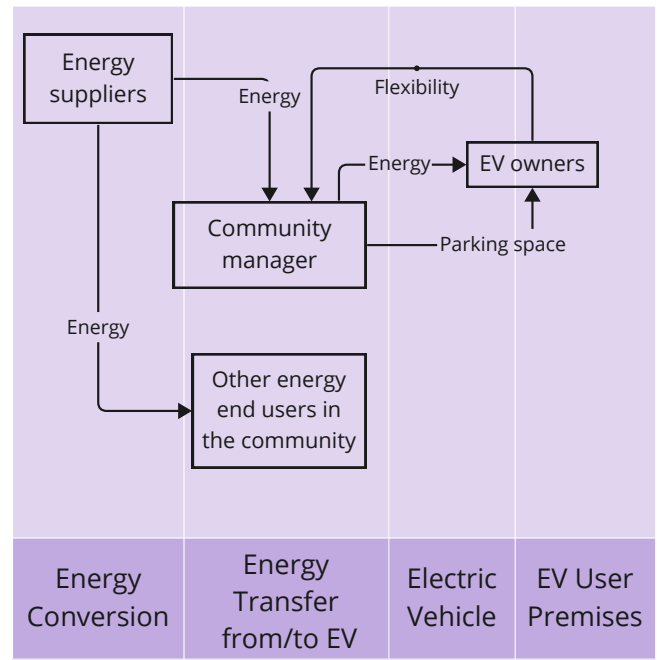

**Figure 9:** Business layer of V2C through the community manager as the facilitator

It's important to note that their participation in flexibility markets may require aggregation, depending on the size of the offers and local regulations. In this situation, aggregators act as an intermediary for the EV owners and flexibility marketplace.

#### 4.1.1. EVs plugged into grid-connected homes

Since a substantial number of EVs are connected to households' electrical systems, they have the potential to play a significant role in addressing challenges at the grid level, provided they can be effectively managed and aggregated. A representation of the component layer of a typical system enabling this participation is illustrated in Figure 11. As is shown in this figure, owners of the EVs plugged into homes can participate in the local flexibility market (through aggregators if it is needed). In this approach the house owners buy energy from the suppliers, and by actively coordinating the charging and discharging process of the EV, they can inform their availability for providing flexibility services to the aggregators or directly to the market system. It should be noted that the aggregation may be needed both for V1G and V2G [85, 86].

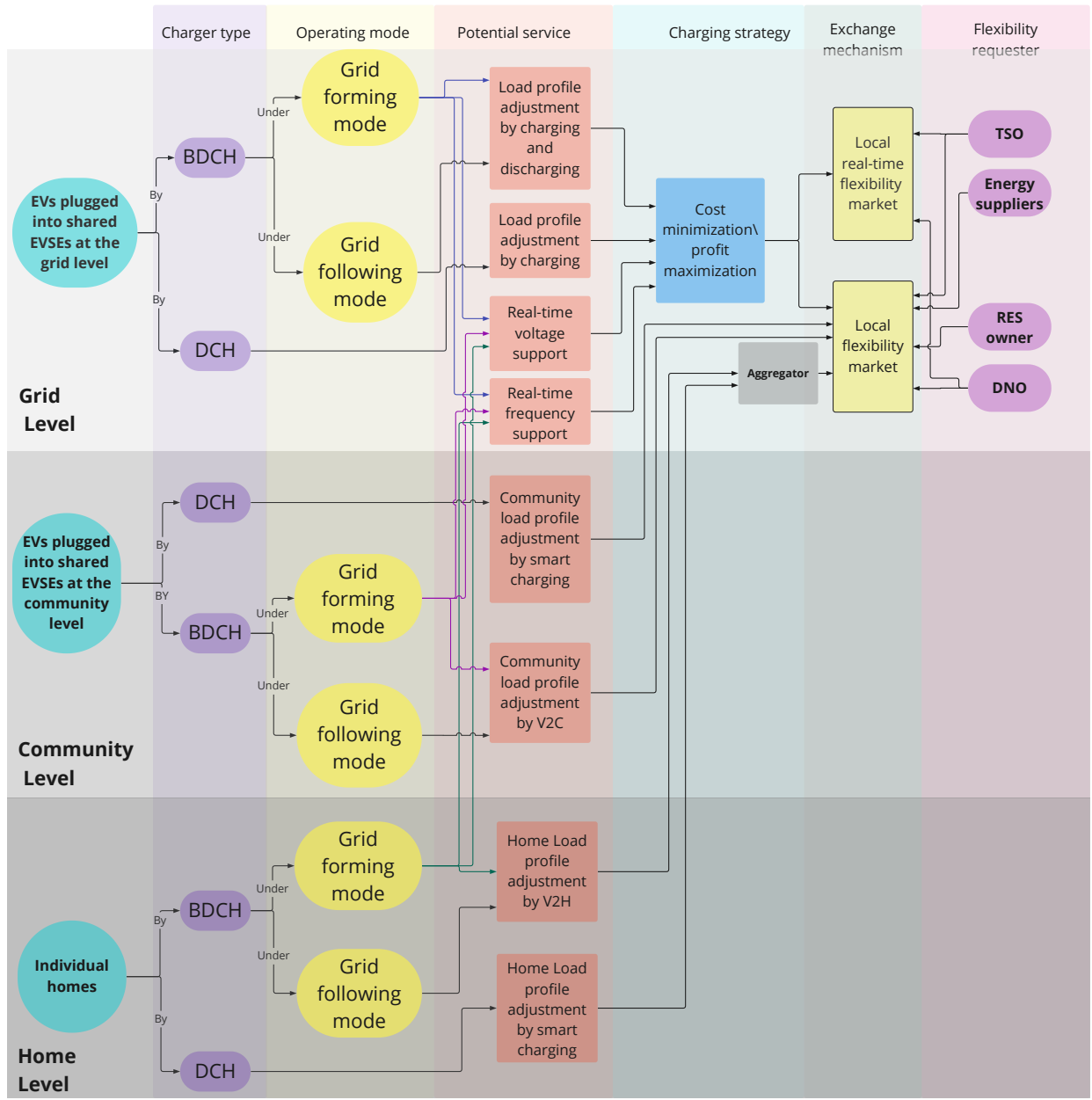

**Figure 10:** Flexibility services at the grid level

### 4.1.2. EVs plugged into grid-connected communities

When multiple EVs are plugged in within a community, there exists a considerable potential to provide flexibility by the community manager, if they are managed properly. The representation of the component layer in Figure **??** outlines the structure facilitating such participation in a typical local flexibility market. The community manager, in coordination with EV owners, can coordinate charging and discharging processes of the EVs. They have the opportunity to convey their combined availability for offering flexibility services to the local flexibility market. This participation can be facilitated by aggregators if the size of the offer is not suitable to be directly presented at the local market.

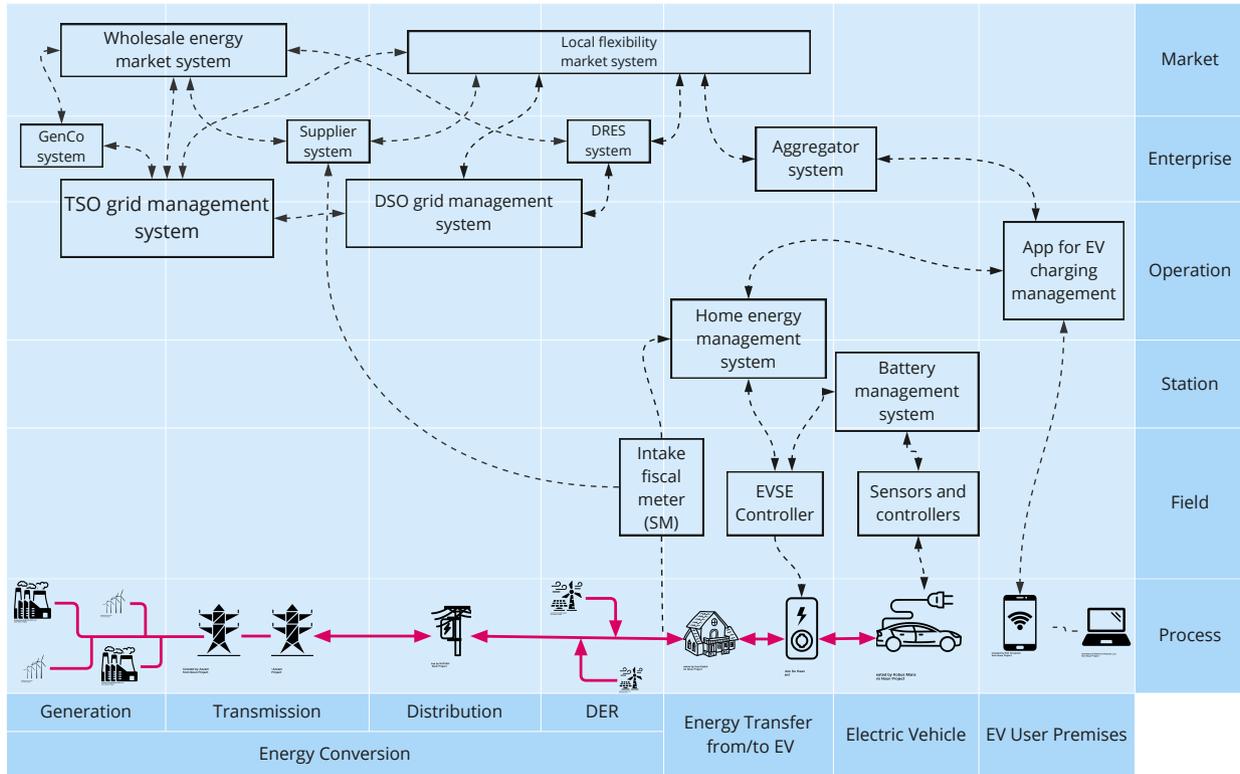

**Figure 11:** Component layer of homes' participation at the grid level

### 4.1.3. EVs connected to public charging stations

Given that the plugged-in duration at public charging stations is typically shorter compared to the plugged-in time at homes and communities, their capacity to provide flexibility is considerably lower than the first two categories. The component layer resembles that of the community case.

## 4.2. Flexibility services at the grid level
### 4.2.1. Service 1: load profile adjustment

This service can be considered as the general version of peak shaving. There are two approaches to adjust energy usage: V1X and V2X. Here X can be homes, communities, and charging stations.

1. Load profile adjustment by V1X

   Under these conditions, without bidirectional charging capability, the EV owner is limited to participating in the load profile adjustment by setting the charging time of the EV. For instance, during peak times when the distribution network faces technical challenges, such as voltage drops more than permissible limits, the EV owner can reduce the household load profile by temporarily stopping the charging process. Another scenario is a distribution network with a substantial share of renewable energy resources. In such a system, the EV can be charged when the energy generated by renewables exceeds demand, allowing the surplus power to be utilized rather than curtailed. This can be referred to as 'demand turn-up'.

2. Load profile adjustment by V2X

   When a charging system at a household or a community is a bidirectional charger, and the EV can also be discharged, the ability of the household owner or the building manager to adjust the load profile increases substantially. In this situation, the charging process of the EVs can be managed to provide some flexibility services and they can also be discharged. This extra flexibility can effectively increase the ability of the homeowners or building managers to provide flexibility.

   This load profile adjustment can be achieved by two different approaches:

   - Implicit load profile adjustment

     In this approach, the charging and discharging of EVs are optimized in response to import and export energy prices to minimize charging costs or maximize discharging profits for EV owners. Therefore, activating this flexibility requires having a time-varying tariff for EVs.

   - Explicit load profile adjustment

     In this approach, EVs have the ability to sell their flexibility as a tradable commodity in the local flexibility marketplace or national-level markets. Thus, this flexibility can be activated either in the presence of a flat or time-varying tariff. So, it can be considered a new revenue stream for EV owners. This dual

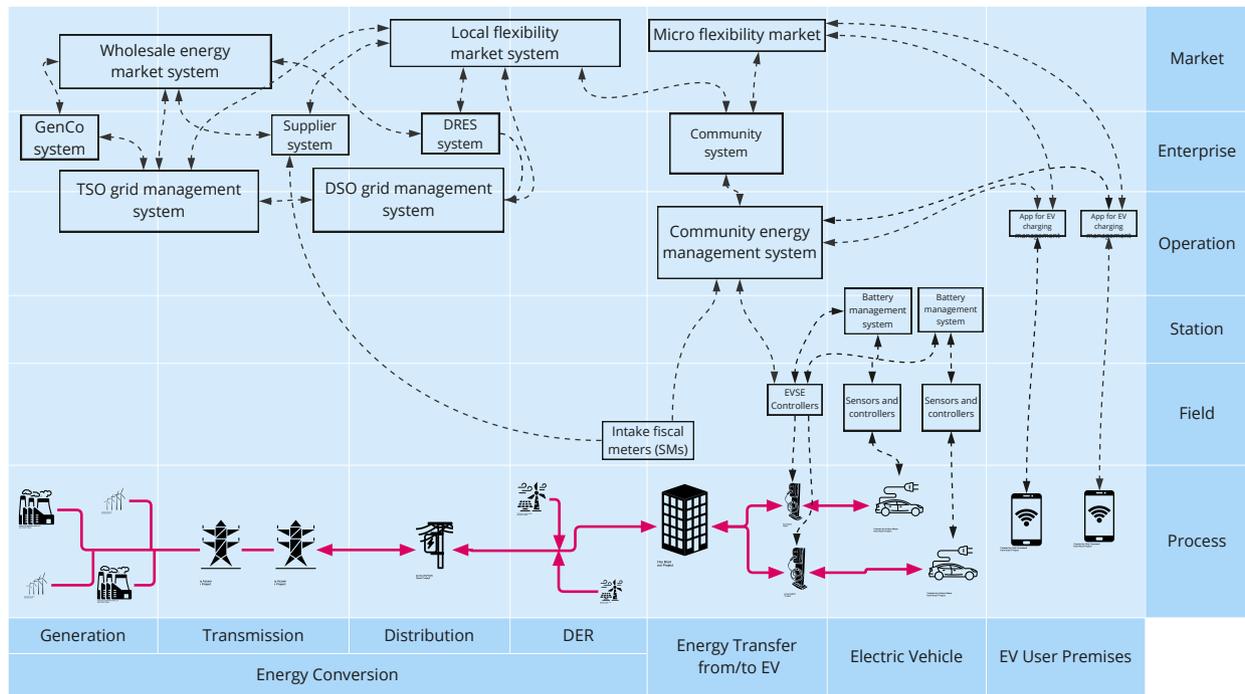

**Figure 12:** Component layer of communities' participation at the grid level

capability enhances the overall value proposition of EVs, making them not only a sustainable mode of transportation but also a more dynamic asset in the power sector.

### 4.2.2. Service 2: real-time voltage and frequency support

This service is achievable solely when the EV charging system in the household or in the community is bi-directional with grid-forming capabilities. To meet these criteria, EV owners must share their availabilities with the aggregator. Subsequently, the aggregator can participate in ancillary service markets, communicating the outcomes to the EV owners. In [87], a unified control system has been proposed for a low-power bidirectional EV charger, allowing it to function as an active power filter and a single-phase four-quadrant static synchronous compensator. The developed charger can regulate both voltage and frequency within the household.

## 4.3. Flexibility requesters at the grid level

### 4.3.1. Distribution System Operators

Distribution system operators need to engage in flexibility markets to procure the necessary flexibility for optimizing grid operations and managing variable renewable energy resources effectively. The management of this flexibility procurement is essential for DNOs/DSOs to address current and anticipated challenges in their distribution network and to achieve their objectives in both operational and planning contexts. Flexibility services that can be provided by EVs to DNOs/DSOs have been categorized in Table 3 and have been explained below.

1. Cost minimization

   EVs have a great opportunity to actively participate in distribution network cost minimization. Distribution network cost minimization can be defined as minimization of the energy losses, improving the operational condition of assets to prolong their lifetimes, deferring reinforcement investment to follow the demand growth, and minimizing the cost associated with maintenance outages [113, 114]. There is also the fact that electricity demand will increase substantially over the coming years with increases in electric heating and mobility. This has the potential to put all transformers under strain and potentially requiring replacement, but careful management of EV charging, which is more flexible than heating, can spread the demand over the day, resulting in the transformer operating within limits and households able to connect their new electrical systems. To achieve this, EVs' charging and discharging schedules should be carefully managed and coordinated. A two-level model for the cooperative management of distribution networks in the presence of multiple charging stations has been presented in [90]. At the distribution system level, the DSO aims to minimize the overall operational cost while determining the optimal energy and reserve prices for transactions with charging station. At the lower level, CSs optimize their individual costs by adjusting bid quantities and setting optimal incentive prices for EV drivers.

2. Peak shaving

**Table 3**
Services that can be provided by EVs to DNOs/DSOs

| Reference | EV technology | Service | Participation approach | EVs activation approach | EV battery degradation modeling |
|---|---|---|---|---|---|
| [88] | V2G | Congestion management | Indirect through virtual energy storage systems | Price-based DR | financially compensated (a linear model) |
| [89] | V1G | Congestion management | Indirect through aggregators | Price-based DR (LMP) | No |
| [90] | V1G | Distribution network cost minimization | Indirect through charging stations | Price-based DR for CSs and incentive-based DR for EVs | No* |
| [91] | V1G | Congestion management | Indirect through aggregators | Price-based DR | No |
| [92] | V2G | Distribution system optimization (social welfare maximization) | Direct | Price-based DR (LMP) | No |
| [93] | V2G | Congestion management and peak shaving | Indirct though charging stations | NA** | No |
| [94] | V2G | Peak shaving | Indirect through aggregators | Incentive-based DR | No |
| [95] | V2G | Distribution network cost minimization | Indirect through parking | Incentive-based DR | No |
| [96] | V1G | Congestion management | Indirect through fleet operator | Price-based DR | No |
| [97] | V2G | Congestion management | Direct | Role-based control*** | No |
| [98] | V2G | Congestion management | Direct | Auction | No |
| [99] | V2G | Congestion management | Indirect through aggregators | Role-based control*** | No |
| [100] | V2G | Distribution network cost minimization | Indirect through parking lots | NA** | No |
| [101] | V2G | Distribution network cost minimization and congestion management | Direct | Price-based DR | No |
| [102] | V1G | Congestion management | Indirect through fleet operator | Price-based DR (LMP) | No |
| [103] | V1G | load balancing | Direct | NA** | No |
| [104] | V1G | Distribution network cost minimization and load balancing | Direct | NA** | No |
| [105] | V2G | load balancing | Indirect through aggregators | Price-based DR | No |
| [106] | V2G | Distribution network cost minimization | Indirect through VPP as the aggregator | Price-based DR (LMP) | No |
| [107] | V1G | Distribution network cost minimization | Indirect through EV charging aggregator | Price-based DR (LMP) | No |
| [108] | V2G | Ancillary services | Indirect through EV aggregator | Price-based DR | No |
| [109] | V1G | Ancillary services | Indirect through fleet operator | NA** | No |
| [110] | V2G | Peak shaving | Indirect through an aggregator | NA** | Yes |
| [111] | V2G | Peak shaving | Direct | Price-based DR | Yes |
| [112] | V2G | Peak shaving | Direct | Price-based DR | Yes |

*Even though an incentive is paid to EV users, there is no guarantee that this covers the degradation cost. This is due to the point that it is calculated by an optimization algorithm to meet Karush-Kuhn–Tucker conditions.
**The financial aspects of the model have not been discussed.
***Rules have been established to regulate the charging and discharging of EVs. However, the financial arrangements between EVs' owners and the DSO/aggregator remain unspecified.

This service aims to reduce the demand on the distribution network during peak hours. This can be achieved by avoiding EV charging during these time slots and, alternatively, by purchasing excess energy from them. Both individual EVs and aggregated groups of EVs can provide such a service. In the literature, it has also been referred to as load-shifting [94], load leveling [115, 116], net-load variance minimization [110, 117, 118], and net-load balancing [111, 119].

3. Congestion management

In the last few years, after the increased penetration of renewable energy resources into distribution systems, distribution networks faced a new challenge. This challenge is due to the fact that renewable generation, such as wind power and solar power, which are not controllable, may produce energy more than the network demand in specific time slots. It should be noted that this has not traditionally been seen as a primary responsibility of the DNO/DSO but new rules require the operator to make compensation payments to generators when curtailment of generation exceeds agreed limits. Therefore, there is a financial imperative to manage generation constraints more actively.

Three primary mechanisms for managing congestion in distribution networks, namely smart tariffs, local markets, and direct control of flexible assets [120]. Typically, these methods are examined independently, neglecting possible interrelations and compromises. In [120], advantages and disadvantages of these mechanisms have been investigated and a comprehensive design framework for managing congestion has been presented. A bi-level- MILP model has been proposed by [88] to activate EVs for congestion management at the distribution network level. At the DSO level of this model, the objective function consists of three components aimed at reducing switching actions, preventing load shedding, and minimizing the ratio of transmitted power to the rated capacity of the lines in order to effectively handle congestion. At the lower level, profit maximization has been used as the objective function of virtual energy storage systems. In [89] a market-driven approach is proposed to enable both EV users and households to engage in a DR program aimed at mitigating network congestion. The methodology employs a Lyapunov optimization framework to adjust EV schedules and household loads. In the proposed approach by this reference, locational marginal prices are used for the calculation of energy cost.

4. Ancillary services

This service can be designed to assist the DNO in maintaining the voltage profile within permissible ranges. To offer such a service, EVs need to be charged and discharged in coordination with the DSO. Whenever the DSO encounters challenges in maintaining the voltage profile within permissible ranges, it should buy this service from EVs or aggregators to inject/absorb energy into/from the network.

Research in this area can be classified into two main categories: those employing EV active power adjustment [109] and those advocating the use of bidirectional chargers to absorb or inject reactive power [121–123]. An operation mode titled vehicle-for-grid has been proposed by [65]. In this operation mode, the EV is exclusively utilized for generating reactive power or functioning as an active power filter, which involves mitigating harmonics in the overall home current.

5. Voltage and load balancing

In low voltage distribution systems, where loads are predominantly single-phase, distributing the load evenly across three phases poses a challenge. Achieving a balanced load profile in all three phases is difficult in such networks. As a result, voltage and current imbalances occur. This imbalance contributes to increased energy losses within the distribution network and a decline in power quality. Coordinated charging and discharging of EVs connected to residential chargers have a promising opportunity to reduce the unbalance indexes and consequently its adverse effects on the network and end users.

### 4.3.2. Transmission System Operator

Currently, at the power grid level, there are several markets for trading different types of services. TSO needs to buy its required services through these markets. Due to the rapid response times, EV and PHEV batteries offer considerable potential for supplying ancillary services, especially frequency regulation services [86]. Aggregators are essential for EVs to engage in electricity markets at the transmission system level, owing to both regulatory and physical constraints. The aggregator consolidates the capacities of numerous EVs and submits their combined capacity as bids into electricity markets [124, 125]. In this situation, the aggregators are responsible for managing the uncertainties in the flexibility provided by EVs. That's why different kinds of uncertainty management approaches, such as fuzzy set theory [124], sequential linear optimization model to minimize the deviations costs [126, 127], Monte Carlo simulations [128], and stochastic programming [129] have been proposed to overcome this challenge. Some of this research has been summarized in Table 4. A V2H aggregator is proposed in [86] to facilitate the participation of EV and PHV owners in the regulation market. To enhance resource efficiency and EV owners' profitability, it's crucial to take into account individual consumption patterns, such as vehicle use and energy usage, when determining the best operational settings for EVs and PHVs.

### 4.3.3. Energy suppliers (retailers)

Electricity retailers are capable of establishing direct connections with both energy consumers and producers. That's why retailers can be considered as a valuable intermediary for managing matters between market operators and demand-side participants [145, 146]. They are required to buy volumes of energy sufficient to meet the consumption needs of the customers they serve, for each half-hourly interval. This procurement process involves various mechanisms, such as entering bilateral agreements with generators and engaging in day-ahead/intraday energy markets for buying and selling electricity, even up to an hour before real-time [147, 148]. Purchasing flexibility from their customers can be viewed as highly valuable for managing their energy portfolio or imbalance cost minimization. This procurement primarily aims for profit maximization. Different aspects of the relationship between retailers and EVs have been discussed in [145, 149–152]. In [151], the optimal energy procurement problem faced by retailers has been studied. In this study, an incentive-based DR program is used to motivate PEV owners to reduce their demand. Different uncertainties associated with PEVs, such as arrival/departure times, daily mileage, and vehicle types have been addressed through an MILP optimization model. Similarly, DR programs have been used in [153, 154] by energy retailers to shape the EVs' loads.

### 4.3.4. Renewable energy resources' owners

The output of renewable energy resources cannot be easily adjusted. This is the main drawback of these resources. In distribution networks with a high penetration of renewable energy, there may arise situations where energy demand is lower than the generated renewable energy. In such cases, curtailment of this excess generation becomes necessary to avoid technical issues in the local distribution system or the broader power system [155]. Usually, DSOs possess the authority to curtail excess renewable power generation as per their installation contract. In such cases, where DSOs are permitted to curtail power in accordance with contractual agreements, owners of renewable energy resources may also

**Table 4**
Services that can be provided by EVs to TSO

| Reference | EV technology | Participation approach | EVs activation approach | EV battery degradation modeling |
|---|---|---|---|---|
| [130–133] | V2G | frequency regulation market | Indirect through aggregators | No |
| [124] | V1G | regulation market and spinning reserves market | Indirect through power aggregators | No |
| [126, 127] | V1G | day-ahead energy market | Indirect through power aggregators | No |
| [129] | V2G | day-ahead energy market, spinning reserve market and regulation market | Indirect through parking lot operator | Yes |
| [134–137] | V2G | supplementary frequency regulation | Indirect through aggregators | No |
| [138] | V2G | day-ahead reserve market | Indirect through power aggregators | No |
| [139] | V2G | Primary frequency control and dynamic voltage support | Indirect through aggregators | No |
| [140] | V1G | day-ahead and real-time energy and frequency regulation markets | Indirect through power aggregators | No |
| [141, 142] | V2G | Primary Frequency Regulation | NM* | Yes |
| [143] | V2G | real-time frequency regulation | Indirect through aggregators | Yes |
| [144] | V2G | frequency regulation | Indirect through aggregators | Yes |
| [17] | V2G | frequency containment reserve and peak shaving | Indirect through aggregators | Yes |

*Not mentioned.

be regarded as flexibility buyers to prevent that curtailment. However, there are regulations in place for DSOs aimed at ensuring they do not surpass contracted curtailment limits. This incentivizes DSOs to function as flexibility buyers, thereby averting curtailment of renewable energy beyond the agreed-upon levels.

A review of the current smart charging methods for integrating renewable energy with EV technology has been provided in [156]. The potential of EVs in integrating renewable energy into distribution systems has been explored in [157–159] from a general perspective. However, since different types of renewable energies possess unique characteristics, leveraging these specific attributes can enhance the performance of their integration with the assistance of EVs. Solar energy follows a relatively consistent daily cycle, with peak output occurring approximately 4 hours before the highest demand for electricity. In contrast, wind energy is characterized by greater variability and unpredictability, influenced by geographical factors [155]. The support of V1G and V2G has been studied in the integration of wind power [160–163], solar power [70, 164] and their hybrid [165, 166] into distribution systems.

### 4.4. Activation of flexibility services at the grid level

Flexibility services trading at the grid level can be conducted using a market mechanism. It should be noted that at this level, the exchange environment in this market is the distribution network. This can pose two main challenges to the market design:

- The first challenge here is the fact that DSOs have two roles: the first, as the operator of the exchange environment of the product; the second, as one of the buyers. Even though the settlement process itself should not increase pressure on the network, the distribution system operator needs to buy flexibility to overcome its challenges and manage its objectives.

- The second challenge is that distribution networks as the exchange environment of this market have a lot of technical constraints that should be managed. Accordingly, the settlement process of this market needs to check the technical constraints of the distribution network. This means that distribution networks' data should be provided to the market operator.

These two challenges need to be overcome at the market design stage. In the future, a significant rise in the number of flexibility providers is expected in the local V2X flexibility market. Nevertheless, the number of flexibility requesters is anticipated to remain comparatively low, even with the market's expansion. This asymmetry, where the number of sellers greatly exceeds the number of buyers, sets the stage for a potential shift toward monopolistic competition. In this type of market dynamic, prices are influenced and determined by the few powerful buyers, for example energy suppliers and distribution system operators.

The business layer at the grid level, when a local flexibility market is in place and the competition has the potential to be pure, has been illustrated in Figure 13. As is evident, all flexibility requesters fulfill their flexibility requirements through the local flexibility market. Given that regulatory frameworks are instrumental in shaping the operation of local flexibility markets and ensuring fair competition among providers and requesters of flexibility, defining these frameworks is a pivotal task in the design of local flexibility markets.

## 5. V2X supporting regulations

Regulation creates a unique environment for companies operating in the energy sector. It plays a crucial role in directing grid infrastructure related businesses such as DSOs(DNOs in UK), TSOs and other players involved in managing the grid asset. Often regulation frameworks extend from grid asset management to market transactions, and in how market platforms are set up, thus, it is crucial to understand the regulatory environment and how it impacts existing but also emerging business outlooks. In the regulated business environment, regulation can be seen to have two faces; firstly, a conservative approach that might be seen to hinder development; secondly, regulation as a key driver in the evolution and modernization of grid infrastructure. Not only does it guide the investment and development of new technologies, but it also ensures the reliability and efficiency of the grid. It is a complex process that requires a balance between innovation, cost-effectiveness, and reliability.

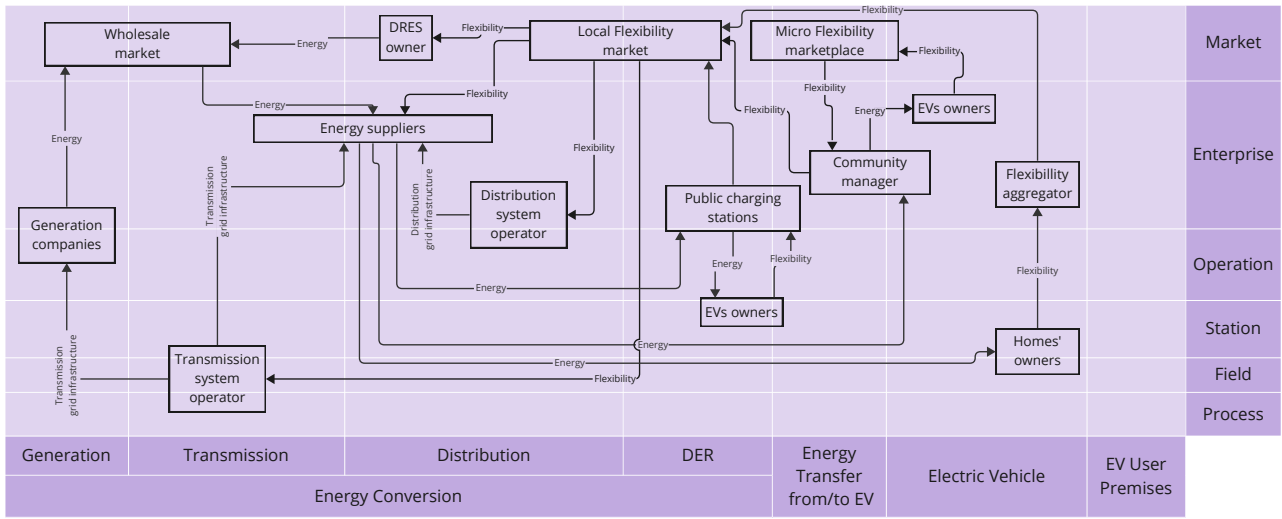

**Figure 13:** Business layer at the grid level when the competition is pure

The EU has a strong push to enable flexibility resources from a variety of sources, including individuals who may be in possession of resources such as building or EVs [167]. Furthermore, the incentive to activate the flexibility of existing infrastructure and mobility is substantial, as if fully enabled, it would contribute to the 2030 EU Green House Gas (GHG) reduction goals 55% [168]. The EU Green Deal Industry Plan aims to support the achievement of the GHG goals by streamlining the regulatory environment. This objective of the plan deals solely with the regulatory environment. In practical terms, this means establishing a simpler, faster, and more predictable framework, ensuring the volume of raw materials needed, and ensuring that users benefit from low-cost renewable energy [167]. To support this work, there are three initiatives: Net-Zero Industry Act, Critical Raw Materials Act, and Reform of the electricity market design. [169], [170], [171] The energy market refomr hast three main objectives: first, a fully integrated European energy market is the most cost-effective means of ensuring a safe, sustainable and affordable energy supply to EU citizens. Second, through common energy market rules and cross-border infrastructure, energy can be produced in one EU country and delivered to consumers in another. Third, increasing competition, improving long-term markets and allowing consumers to choose energy suppliers keep prices under control.

Regulation can be a powerful tool to guide the development of the energy system and energy markets, but it is not without challenges. The challenges have been pointed out by, for example, article [172], which states that flexibility in the planning and operation of electricity networks is a significant change from the current situation in which DSOs have full control over their network assets to a new system in which DSOs should rely on third-party flexibility services. In this new regime, the safe operation of the network relies on more actors, more ICT systems, and more links (direct and indirect) between the network operator and the network user. The uncertainty of resource availability plays a key role in the emergence of new sources of flexibility, such as EVs capable of V2G. Trial projects have shown that the resource portfolio needs to be large enough to address doubts about availability [38]. In addition to resource availability, V2G is currently regarded a similar resource as any other storage. Exposing the V2G application to double taxation, meaning that taxes are paid when charging the battery, but also when discharging. This is a difficult issue when the V2X application is energy intensive, for example when taking part in the SPOT markets, the value stream is heavily impacted by double taxation [173]. This is an issue that can only be solved by the regulatory entity, such as the EU Commission or national regulatory entities.

In summary, the regulatory framework has a strong impact on guidance in the European market area, but it alone cannot enable or launch ecosystems that enable novel flexibility resources such as V2G. The V2G has many identified barriers, in technical, market, and social aspects. The ecosystem needs a clear indication of the solid value streams associated with the regulatory environment that support development.

## 6. Trial projects

In this section, V2X-related pilot projects in the EU and UK are elaborated on to illuminate the lessons learned from them and to provide a more realistic view of the V2X opportunities and challenges.

**Table 5**
Summary of the UK's current markets available at the power system level related to frequency response, reserve and flexibility

| | Static Firm Frequency Response (FFR) | Balancing Mechanism | Short term operating reserve (STOR) | Fast reserve | Demand Flexibility Service |
|---|---|---|---|---|---|
| time to respond | 30 seconds | Defined by provider (1 minute-89 minutes) | 20 minutes | 2 minutes | 7.5 hours |
| Delivery duration (minutes) | 30 minutes | 15 minutes | Minimum 2 hours | 15 minutes | 30 minutes |
| Minimum volume of the offer (MW) | 1 | 1 | 3 | 25 | 1 |
| Procurement window | Daily auction | 1 hour ahead of real-time | Day-ahead auction | Real-time | Day-ahead |
| Metering requirement | Real-time active power and real-time frequency for at a rate of 1Hz | active power at a rate of 1Hz | active power at a rate of 1Hz for BM units for non-BM units every 15 seconds | active power at a rate of 1Hz for BM units for non-BM units every 15 seconds | Half-hourly active power |

**Table 6**
Summary of V2G trials from the point of view of the flexibility services [168, 175, 176]

| | Powerloop | Sciurus |
|---|---|---|
| Region | UK | UK |
| Provided services | Participation in balancing mechanism | Participation in Dynamic Containment(DC) and Firm Frequency Response (FFR) |
| Explicitly vs. implicitly of the provided services | Explicit and implicit | Explicit and implicit |
| Responsibility for charging and discharging of EVs | Octopus Energy Group's Kraken platform as the aggregator | Kaluza's Intelligent Energy Platform |
| Participants profits | £840 per year comparing to unmanaged charging and £180 per year comparing to smart charging | £410 per year comparing to unmanaged charging and £172 per year comparing to smart charging * |
| Number of EVs | 135 Nissan LEAF (95% of the EVs were the 40 kWh version model, and the rest were the 62 kWh model) | 325 Nissan EVs |
| Note on metering | The measurements, including active power and SOC, are gathered by the charger at a frequency of 0.1 Hz. Thus, the current metering entry requirement has not been fulfilled. Instead, a combination methodology has been used to achieve the required 1 Hz granularity. | The participation in DC and FFR has been theoretically studied; however, the necessary entry metering requirements have not yet been fulfilled. |

* These values are the averages as stated in [177]. For example, for a 40kWh battery, smart charging tariff optimization, V2G tariff optimization, V2G + FFR, and V2G + FFR + DC can result in £120, £340, £513, and £725, respectively [176].

Table 5 provides a broad overview of UK markets accessible to providers seeking entry with energy-limited assets, like EV charge points. They outline key parameters for each market, aiding in the assessment of the asset's suitability. While not an exhaustive list of ESO markets, these tables concentrate on those deemed most relevant to this asset type [147]. As can be seen from the table, the balancing mechanism, short-term operating reserve, and fast reserve need an active power measurement with a rate of 1 Hz, which is high compared to the measurement capabilities of current smart meters. It should be noted that other UK's current markets available at the power system level related to dynamic frequency services such as dynamic containment, dynamic moderation, and dynamic regulation need real-time active power at a rate of 1 Hz for delivery check and measurements of active power and frequency at a rate of 20 Hz for performance monitoring. This seems far beyond the capabilities of smart meters. For instance, smart meters in Great Britain can measure at 10-second intervals with added equipment, yet even this is inadequate for participation in such markets [174].

A summary of two of the most promising V2G trials from the point of view of the flexibility services has been presented in Table 6. The first one entitled "Powerloop" has been delivered by Octopus Energy and Octopus Electric

Vehicles, bi-directional V2G chargers (Wallbox Quasar) have been installed [168]. In this project, Octopus Energy and National Grid ESO used a test environment of the balancing mechanism to demonstrate that EVs can interact with the ESO and support power system balancing. In this trial, the Octopus Energy Group's Kraken platform was responsible for managing EVs' charging and discharging schedules. It was working in the background to align the customer's schedule with grid signals to offer flexibility as a service [168]. Upon receiving dispatch signals from the system operator, this platform promptly adjusted the schedules of all EVs to deliver an aggregated response to the system operator's signal while ensuring that all EV owners' requirements would be satisfied. The second one is titled "SCIURUS". The main objective of this V2G trial was to "Validate the technical and commercial potential for a domestic V2G charging solution" [176].

## 7. Opportunities for monetizing V2X

Approaches in which EVs can generate profits through V2X opportunities are explained in this section. These approaches can be classified into two categories: 1) explicit flexibility through direct participation and 2) implicit flexibility through indirect participation. Direct participation involves the active involvement of individual EVs or aggregated participation as a group in various flexibility markets. In this manner, they can profit based on their contributions and the outcomes of market settlements. In indirect participation, EVs can generate profits without engaging directly in flexibility markets. For example, they can charge their EVs in off-peak when the energy prices are low and then utilize a portion of that stored energy during peak hours when energy prices are higher.

### 7.1. Current real-world examples for explicit flexibility

In this category, there are some under-development and completed platforms and trials as follows:

#### 7.1.1. Octopus power pack (UK)

Octopus Power Pack manages V2G charging and exporting schedules. Users can plug in their cars and specify their desired charge levels. The system will then automatically schedule the vehicle to charge during times of greenest energy availability and subsequently export surplus energy to support the local grid when needed. The end users need to have a Wallbox Quasar 1 V2G charger and a compatible EV (Nissan Leaf, Nissan e-NV200, Mitsubishi Outlander PHEV) to be able to use this plan. Also, G99 application needs to be filled to be able to export power to the grid. It demonstrates that the connection adheres to Engineering Recommendation G99, as issued by the Electricity Networks Association[2].

#### 7.1.2. PicloMax (UK)

This platform is under development by Piclo It will provide streamlined access to all electricity markets (long-term, situational, day-ahead and intraday) to enable flexibility providers such as EVs, batteries, and renewables to maximize their asset revenue [3].

### 7.2. Current real-world examples for implicit flexibility

In the UK there are multiple DR programmes that can facilitate the participation of EV owners. Since a considerable number of EVs are parked in residential areas during these hours, both V2G and even V1G technologies have the potential to significantly increase participation rates. "Octopus Saving Sessions" and "Demand Side Response" are examples of these DR programs in the UK. The first one is proposed by Octopus Energy. In the program, end users are paid to decrease their electricity consumption during peak times. Those with EVs can adjust their charging schedules to reduce their demand, or alternatively, they can leverage the V2H option, using the energy stored in their EVs' batteries to further reduce their demand. The second one has been introduced by E.ON company. It is a strategic approach enabling energy consumers to reduce their electricity usage during peak demand periods. Similar DR programs are available in the EU. A practical V2H platform to facilitate the participation of end users in DR programs has been developed in [178]. In this study a commercialized EV with CHAdeMO port has been used for experimental setup.

## 8. Conclusion and future directions

This paper identified three levels for practical V2X flexibility services: home level, community level, and grid level.

- At the home level, there is no need for any exchange mechanism since the owner of the EV and the flexibility requester are the same.

- At the community level, the community manager as well as EV owners can be considered as the flexibility requesters. Since there may be some conflict of interest between the EV owners, as the flexibility providers, and flexibility requesters, an exchange mechanism will be necessary.

- At the grid level, all the EVs that are supplied from the distribution system, including EVs plugged in homes or communities or public charging stations, have the potential to be a flexibility provider. On the other side, DNOs/DSOs, energy suppliers, aggregators, TSOs and renewable energy providers can harvest some opportunities by requesting flexibility. Therefore, at the grid level, there may be multiple flexibility requesters and a lot of small-scale flexibility providers. That's why having a fair and transparent exchange market is necessary at this level.

There is extensive research focused on the participation of EVs in different markets at high-voltage levels of power systems, while there is comparatively little dedicated to distribution systems. The main challenges for EVs participating in these markets at the transmission level are the entry requirements. The most challenging entry requirements for EVs include the minimum value of the offer, response time, and metering requirements.

- Minimum value of the offer: Currently, the minimum value of the offer at the transmission level is 1 MW per geographical point. Although this challenge seems easy to overcome through aggregation, it still poses a significant hurdle.

- Response time: For those markets that need very fast response time, such as dynamic containment, and dynamic moderation in the GB, where the initiation time is as short as 1 second and the service must be fully delivered within the subsequent 0.5 seconds, or even dynamic regulation in GB where the initiation time is 10 seconds and the service needs to be fully delivered within the subsequent 2 seconds, the response time poses a significant challenge for EVs. On the other hand, for markets with slower response requirements like static firm frequency response, where the time to respond is 30 seconds, while still

---

challenging for EVs, it appears achievable through automated responses from already plugged-in EVs. In markets with slower response requirements such as the balancing mechanism and short-term operating reserve in GB, where the time to respond is on the order of 10 minutes, it is easier for EVs to provide their flexibility services.

- Metering requirements: For instance, involvement in frequency regulation and reserve markets necessitates active power and frequency measurements at a rate of 1 Hz. Given that this measurement rate exceeds that of smart meters, it presents a significant challenge for the widespread participation of EVs in these markets.

Considering all three challenging entry requirements for EV participation in transmission-level markets, it's evident that overcoming technical and regulatory challenges is necessary. While EVs may seem suitable for certain markets such as the balancing mechanism and short-term operating reserve in GB in terms of response time, the requirement for metering at a rate of 1 Hz presents a significant obstacle given the limitations of the current smart meter infrastructure in GB. Addressing these challenges could unfold in three potential directions:

1. Enhancing measurement rates: One approach involves increasing measurement rates at the charger level to meet the requirements of frequency regulation and reserve markets. This improvement could occur by installing more precise meters directly at the chargers as the EV's sub-meter.

2. Revising market entry requirements: Another approach is to simplify the entry requirements of these markets to accommodate the capabilities and limitations of EVs. This approach seems challenging since it can affect the security of the whole power sector.

3. Establishing flexibility markets: A third and perhaps more promising option is the creation of new markets specifically tailored to leverage the flexibility of EVs.

Considering these potential directions, the establishment of new markets specifically for trading the flexibility services that can be provided by EVs seems to be the most viable solution. By designing markets that align with the capabilities of EVs while addressing their technical constraints, it becomes possible to unlock their full potential as grid resources and enhance overall system flexibility.

Regarding implicitly and explicitly of the flexibility services that can be provided by EVs, it is important to note that both types offer distinct advantages and opportunities for integration into the energy market. The advantages of explicit flexibility services are greater than those of implicit services. Although explicit flexibility services have some advantages compared to implicit flexibility services, their implementation presents more challenges. These advantages and challenges are as follows:

- Explicit flexibility services can be provided to different actors in the energy ecosystem such as retailers, and owners of uncertain intermittent energy resources while implicit flexibility resources can be harvested by those who have the power to change the energy tariffs. On the other hand, to enable this advantage a flexibility marketplace needs to be designed to provide the opportunity for all the flexibility requesters to bid for their required flexibility. To do that, the market operator's role should be regulated. Also, to be able to trade the small-scale flexibility services that can be proved by EVs in this marketplace some other roles such as flexibility aggregator need to be defined and regulated.

- Explicit flexibility services provide more controllability for the flexibility requester. For example, a DSO can buy this type of flexibility service from a flexibility market to decrease the load of a transformer in a substation. On the other hand, to be able to overcome this overload problem by implicit flexibility services locational marginal prices may have to be applied. By this approach, reducing the load of this transformer by implicit flexibility services needs tariff adjustment, which has two disadvantages. First, this tariff adjustment may not fall into the power of the DSO; and second, increasing the energy price for the end users supplied by a transformer to decrease its load seems unfair to those end users compared to other end users supplied by other transformers.

- Another challenge of explicit flexibility services is ensuring delivery confirmation. This challenge can be addressed by assigning the responsibility of directly controlling the chargers to the flexibility requesters.

The technical and regulatory challenges to be overcome to activate explicit flexibility services will delay mass adoption. These challenges encompass various aspects, from metering requirements to addressing regulatory frameworks and market mechanisms. Moreover, the complexity of coordinating and verifying the delivery of explicit flexibility services poses additional obstacles. In contrast, implicit flexibility, while offering fewer advantages compared to its explicit counterpart, has a higher likelihood of widespread adoption in the near future. Its inherent simplicity and compatibility with current grid operations make it a more feasible option for immediate deployment. Therefore, although explicit flexibility services offer more benefits, their widespread implementation depends on overcoming major technical, regulatory, and operational hurdles. Meanwhile, implicit flexibility is ready to address current grid management needs and prepare for future demand-side management innovations.

## Acknowledgement

The authors would like to thank the funding for DriVe2X research and innovation project from the European Commission and the UKRI, with grant numbers 101056934 and 10055673, respectively.

The authors would also like to extend their thanks to all members of Working Package 3 of the DriVe2X project, with special acknowledgment to Dr. Ander Zubiria Gómez from TECNALIA Research & Innovation, Durango, Spain, for his valuable feedback on the initial version of this work.